# LanTu: Dynamics-Enhanced Deep Learning for Eddy-Resolving Ocean Forecasting

Qingyu Zheng[a], Qi Shao[b], Guijun Han[a], Wei Li[a,*], Hong Li[a,*], Xuan Wang[a]

[a] *Tianjin Key Laboratory for Marine Environmental Research and Service, School of Marine Science and Technology, Tianjin University, Tianjin 300072, China*

[b] *Fujian Key Laboratory on Conservation and Sustainable Utilization of Marine Biodiversity, Minjiang University, Fuzhou 350108, China*

* Corresponding author.

*E-mail address:* liwei1978@tju.edu.cn (W. Li), hongli@tju.edu.cn (H. Li).

## Abstract

Mesoscale eddies dominate the spatiotemporal multiscale variability of the ocean, and their impact on the energy cascade of the global ocean cannot be ignored. Eddy-resolving ocean forecasting is providing more reliable protection for fisheries and navigational safety, but also presents significant scientific challenges and high computational costs for traditional numerical models. Artificial intelligence (AI)-based weather and ocean forecasting systems are becoming powerful tools that balance forecast performance with computational efficiency. However, the complex multiscale features in the ocean dynamical system make AI models still face many challenges in mesoscale eddy forecasting (especially regional modelling). Here, we develop LanTu, a regional eddy-resolving ocean forecasting system based on dynamics-enhanced deep learning. We incorporate cross-scale interactions into LanTu and construct multiscale physical constraint for optimising LanTu guided by knowledge of eddy dynamics in order to improve the forecasting skill of LanTu for mesoscale evolution. The results show that LanTu outperforms the existing advanced operational numerical ocean forecasting system (NOFS) and AI-based ocean forecasting system (AI-OFS) in temperature, salinity, sea level anomaly and current prediction, with a lead time of more than 10 days. Our study highlights that dynamics-enhanced deep learning (LanTu) can be a powerful paradigm for eddy-resolving ocean forecasting.

# 1. Introduction

The vast oceans connect different continents, breaking down the barriers of distance and facilitating the thriving development of global trade (*1–4*). The dynamic changes in the marine environment are becoming increasingly significant in their impact on human society and economic activities (*5–7*). Spatiotemporal multiscale processes and complex ocean-atmosphere interactions have a direct impact on the safety of maritime activities, making it necessary to improve the understanding and forecasting of ocean evolution. In the global ocean circulation system, mesoscale eddies (50~330 km), as a crucial source of ocean kinetic energy, not only shaping thermohaline structure (*8, 9*), but also influencing weather and climate through interactions with the atmosphere (*10, 11*). Existing studies have shown that the vertical mixing induced by mesoscale eddies controls local circulation structures and the distribution of temperature and salinity, significantly impacting fishery resources (*12, 13*). The short-term variations of the eddies, along with their interactions with the atmosphere, can trigger extreme sea conditions and weather events, posing a threat to shipping safety and efficiency (*14, 15*). Therefore, accurately forecasting the dynamic behavior of mesoscale eddies and the three-dimensional state of the ocean under their influence is crucial. In recent decades, numerical ocean forecasting systems (NOFS), based on physical frameworks and numerical methods, have been widely used in global or regional operational ocean forecasting. With the support of supercomputing resources, data assimilation technologies and ocean observation systems, the NOFS have rapidly developed (*16, 17*). However, existing NOFS still faces many challenges in eddy-resolving ocean forecasting.

Firstly, the evolution of mesoscale eddies involves complex nonlinear dynamics (chaotic characteristics), which places higher demands on the rationality and integrity of the physical framework within NOFS. Currently, generic global ocean circulation models have large uncertainties in forecasting mesoscale processes. The uncertainty mainly arises from initial conditions, boundary forcing, numerical computations and parameterization schemes (*18*), which can lead to significant deviations in forecast results. In addition, the evolution of mesoscale eddies involves deformation, merging or splitting (*9*). A numerical grid (spatial resolution) of at least 10 kilometres is usually required for NOFS if these processes are to be captured. In operational forecasting, large-scale data assimilation and eddy-resolving ocean simulations consume vast computational resources, which can be challenging for less economically developed regions, especially in ensuring navigational safety.

In recent years, with the advancement of high-quality data products (observations, simulations and reanalysis) for the atmosphere and ocean, data-driven methods based on Artificial Intelligence (AI) have been successfully applied to global weather (*19–23*) and short-term ocean forecasting (*24–27*). These AI-based forecasting systems demonstrate competitive performance while significantly reducing the computational burden compared to traditional numerical methods. Most existing AI-based ocean forecasting systems (AI-OFS) employ global eddy-resolving ocean reanalyses to complete model training. However, there are still large gaps in the forecasting performance of these globally available AI-OFS for different regions, even with the same model backbone and training data. In addition, the local smoothing of the global AI-OFS weakens the variability of mesoscale eddies, which makes it difficult for the eddy signals in the initial conditions to persist beyond two week, even if they enter the AI model (blurring effect). The aforementioned issues lead to potential risks in the application of global AI-OFS for regional eddy-resolving ocean forecasting.

In this study, we present LanTu, a customised ocean forecasting system based on artificial intelligence for region-specific eddy-resolving forecasting of upper ocean (0~643 m). LanTu, designed for regional ocean modeling, incorporates the cross-scale interactions between the ocean and atmosphere into the model inference. In order to improve the ability of the LanTu model to capture and sustain mesoscale eddies, a dynamics-enhanced constraint guided by knowledge of eddy dynamics is designed for targeted optimisation of LanTu. The evaluation shows that the performance of LanTu significantly outperforms several operational systems, including NOFS and AI-OFS. It extends the lead time of skillful forecasts by 2 to 4 weeks, covering variables such as ocean currents, temperature, salinity and sea level anomaly (SLA). The forecasting performance of LanTu for mesoscale eddy dynamics (such as merging and splitting) is better than XiHe (a global AI-OFS), in particular the ability to capture 3D eddy structure. Overall, our contribution to this work can be summarised as follows:

- We propose an AI-OFS for regional eddy-resolving ocean forecasting, called LanTu, which aims to improve the learning and forecasting skills of AI model for mesoscale eddy dynamics.

● The dynamics-enhanced multiscale constraint effectively improves the forecasting performance of LanTu, making it superior to the advanced NOFS and AI-OFS. This improvement extends the lead time for skillful forecasts of ocean states by 2 to 4 weeks.

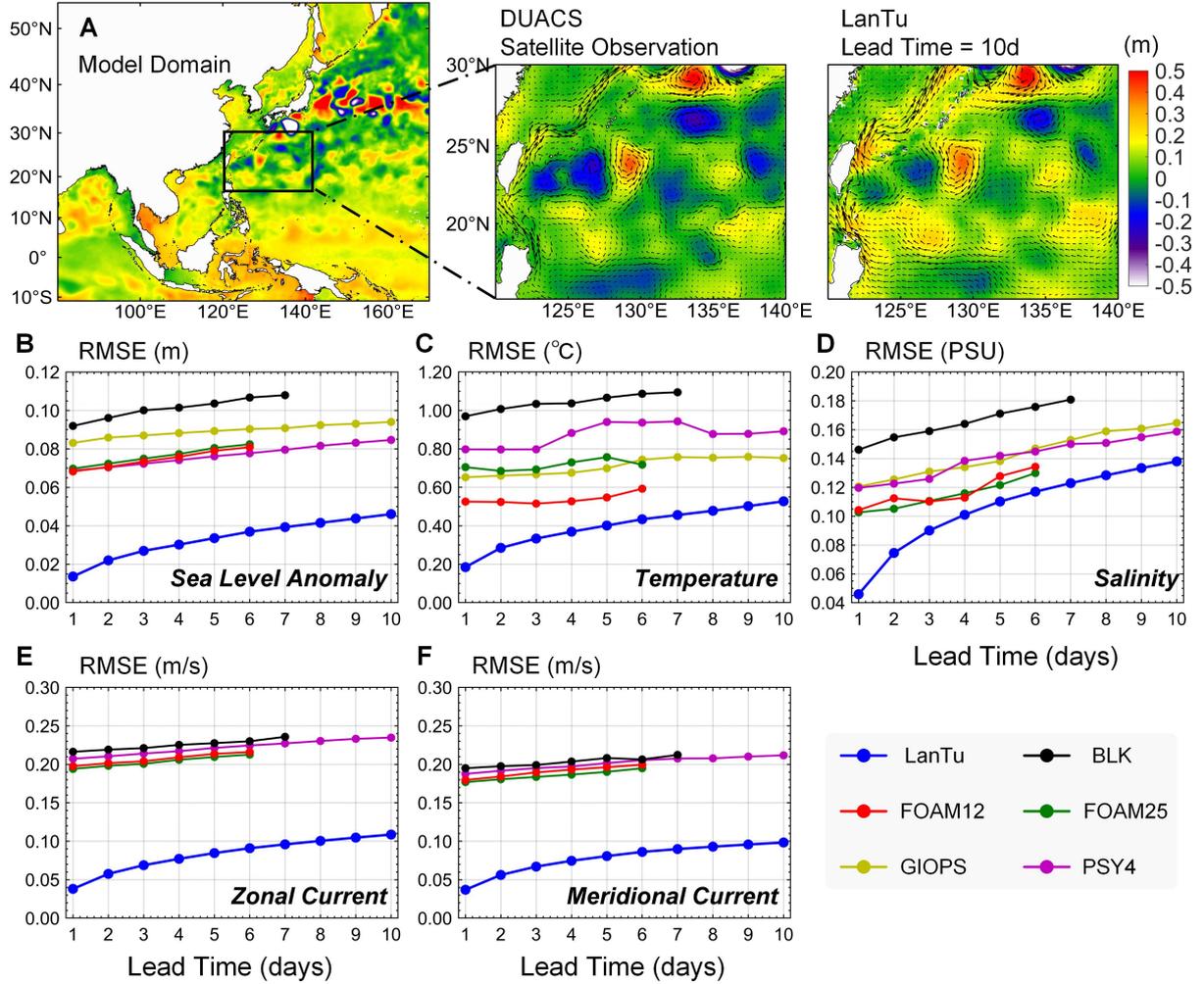

**Fig. 1.** Overall performance evaluation of LanTu. (**A**) A sample of LanTu forecast with a 10-day lead time for SLA on January 15, 2022, referenced to DUACS satellite observation. (**B-F**) Comparison of LanTu with the NOFS from the IV-TT in terms of RMSE, representing SLA, temperature, salinity, zonal current and meridional current, respectively.

## 2. Results

The LanTu model for eddy-resolving ocean forecasts is trained on the current state-of-the-art global ocean reanalysis (GLORYS). LanTu adopts a non-autoregressive modeling strategy to predict the 3D ocean state variables (temperature, salinity and current) and sea level anomaly (SLA) at a daily 1/12° resolution at the target time. LanTu employs a vision transformer (VIT) as the backbone to extract the ocean-atmosphere interactive features and employs dynamics-enhanced multiscale constraint to guide the model in capturing strongly nonlinear evolution. For more details on the model architecture, datasets, model training and evaluation process, see the Materials and Methods section.

## 2.1. Assessment of the LanTu Model

To evaluate the forecasting performance of LanTu, we employ the advanced GODAE Ocean View Inter-comparison and Validation Task Team (IV-TT) Class 4 framework (*28*) for comparison and validation. IV-TT provides observations of currents, temperature profiles, salinity profiles and SLA from drifting buoys, Argo and satellite remote sensing. We employ five different NOFS (Table S1) published by IV-TT as numerical forecast benchmarks for comparison with LanTu and extract assessment data according to the LanTu model domain (Fig. 1A). The evaluation period spans from 2021 to 2023 and the results are shown in Fig. 1. Overall, LanTu exhibits a lower root mean square error (RMSE) across all ocean variables compared to the numerical forecast benchmark in IV-TT (Fig. 1B to F). When the forecast lead time is 6 days, the RMSE of temperature and salinity forecasts for LanTu is reduced by 46.78% and 20.08% on average compared to IV-TT. The performance advantage of LanTu also persists on all variables when the forecast lead time is 10 days.

In order to gain more insight into the forecast errors at different depths, we compare LanTu and IV-TT temperature and salinity profiles in the upper 643 m of the ocean. Both for the temperature profile and the salinity profile, the overall RMSE of LanTu is lower than that of IV-TT (Fig. 2 and Fig. S1). When the forecast lead time is 10 days, the vertical average of the RMSE for the temperature (salinity) profiles forecasted by LanTu is 35.93% (14.81%) lower than that of IV-TT. It can be found that the forecast errors of LanTu and IV-TT varies with depth, where the temperature RMSE shows a regular vertical structure (Fig. 2A to D). It is well known that the ocean is not vertically uniform and thermoclines typically have greater forecast errors (*29*). The temperature error of IV-TT at water depths of about 100-200 m (near the thermocline) is significantly higher than at other depths, with an increase of about 1°C compared to the surface RMSE. The RMSE increase in the thermocline for LanTu is less than 0.5°C, which supports the superior performance of LanTu over NOFS in 10-day short-term ocean forecasts. It is important to note that 10 days is the maximum lead time provided by IV-TT, but it is not the upper limit of forecasting for NOFS or AI-OFS. The profile RMSEs of temperature and salinity (Fig. S1) shows that the 30-day forecasts of LanTu are similar to or even better than the 10-day forecasts of IV-TT. Therefore, the data-driven LanTu outperforms mainstream NOFS benchmarks such as PSY4, GIOPS, BLK and FOAM in ocean forecasting for at least a 10-day lead time.

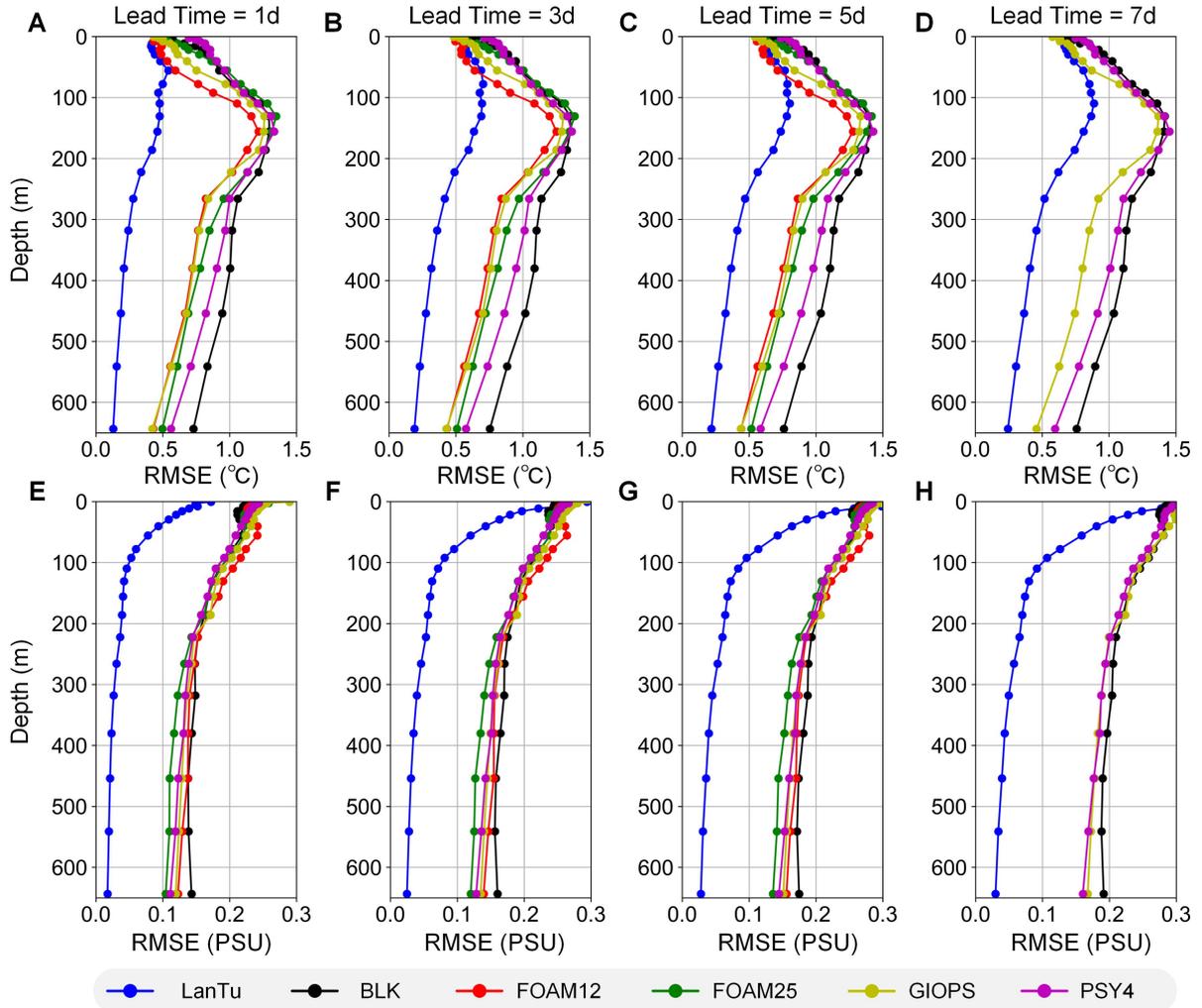

**Fig. 2.** Comparison of RMSE for (**A** to **D**) temperature profile and (**E** to **H**) salinity profile forecasts between LanTu and the NOFS from the IV-TT. Forecast lead times of 1-7 days. The maximum depth is 643m.

In order to quantify the forecasting performance of the LanTu model with a lead time greater than 10 days, we employ the persistence prediction as a reference benchmark, which is used to calculate the persistence skill score (PSS). At the same time, we have developed an autoregressive version of LanTu (LTAR) as the AI baseline model. LTAR needs to predict atmosphere and ocean state variables, using the output from the previous step as input for the next step. In the comparison, LTAR and LanTu have the same model structure and initial field, which can directly reflect the differences between the different modelling strategies. It is worth noting that LTAR has not undergone any fine-tuning strategies (*27*) to mitigate cumulative error caused by autoregressive forecasting.

The PSS of LanTu is consistently greater than 0 and shows a positive trend over the 10-30 days lead time, which can indicate that LanTu can provide more accurate forecasts than persistence for more than 2 weeks (Fig. S2). We find that LTAR fails to outperform persistence in forecast lead times above 10 days for all ocean variables. In particular, the negative slope exhibited by LTAR is significantly steeper than the positive slope of LanTu, which is due to the faster accumulation of errors caused by autoregressive strategy. Therefore, the LanTu can mitigate the cumulative error without adding extra burden.

Unlike RMSE, which quantifies the distance between forecasts and observations, the spatial anomaly correlation coefficient (ACC) can be employed to evaluate the forecasting skill of the LanTu. When forecasting 30 days in advance, the ACC of LanTu for temperature, salinity and SLA is greater than 0.6 . Persistent forecasting retains multiscale information from the initial field. As the lead time increases, small-scale signals gradually lose their effectiveness, while large-scale signals become dominant. Although the forecasting skill for ocean currents is slightly lower than other variables, LanTu can still provide usable forecasts for about two weeks, consistently outperforming persistence (Fig. S3).

Fig. 1A shows a comparison of the local details between the 10-day SLA forecast from LanTu and satellite observations. It can be observed that LanTu successfully reproduces the spatial pattern of mesoscale eddies in the observations, particularly the relative positions and influence ranges of cyclonic and anticyclonic eddies (*30*). To explore the interpretability of LanTu for capturing dynamics features, we select five (Cases 1~5) zonal sampling bands (Table S2) to plot the Hovmöller diagram (Fig. S4), representing different ocean dynamics environments. We found that LanTu successfully captures the dynamical patterns in different regions. Strong westward propagating fluctuations and periodic signals are the main features in Case 1 (Fig. S4B). After crossing the Luzon Strait, the dynamical pattern changes. Similar results are also reflected in Case2 (Fig. S4C). The consistency between LanTu and satellite observations further shows the physical reliability and interpretability of the LanTu forecasts..

## 2.2. Enhanced Forecasting of Mesoscale Eddy Dynamics

In this subsection, we compare LanTu with a well-established AI-OFS, namely XiHe. LanTu and XiHe are both trained on GLORYS and ERA5 reanalysis data, focusing on ocean processes in the upper 643m and aiming to provide eddy-resolving (1/12°) ocean forecasts. It is worth noting that LanTu and XiHe employ the same non-autoregressive modelling strategy, with the main difference being the design of the model architecture and constraint.

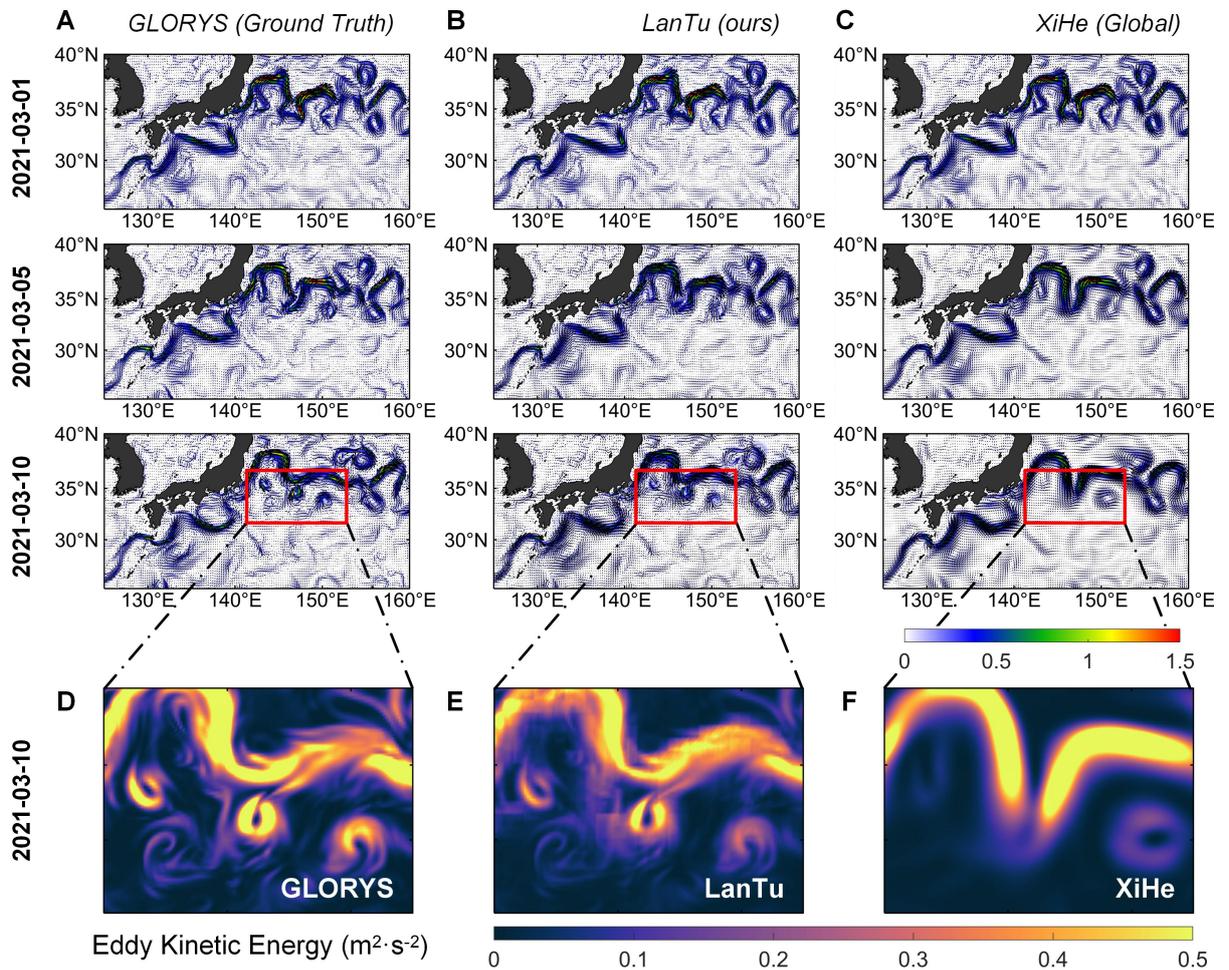

**Fig. 3.** Comparison of the dynamical processes of mesoscale eddies in the AI-OFS forecast. (**A**) Ground truth (GLORYS), (**B**) LanTu, (**C**) XiHe, showing the evolution of sea surface current and EKE. The forecast starts on February 28, 2021. The three rows from top to bottom show the results for March 1, 2021, March 5, 2021 and March 10, 2021. (**D** to **F**) Zoom-in of the local EKE forecasted 10 days in advance.

We find that the variability of mesoscale eddies is severely underestimated by XiHe, but LanTu can effectively forecast the evolution of mesoscale eddies (Fig. 3). When forecasting 10 days in advance, LanTu can accurately capture the shedding of three eddies (Fig. 3A and B), while XiHe predicts only one(Fig. 3C). In addition, the variability of mesoscale eddies in XiHe is strongly suppressed, and the eddy kinetic energy (EKE) is smoother. The overall EKE details of the LanTu forecasts are very similar to those of GLORYS, although there is also smoothing and suppression.

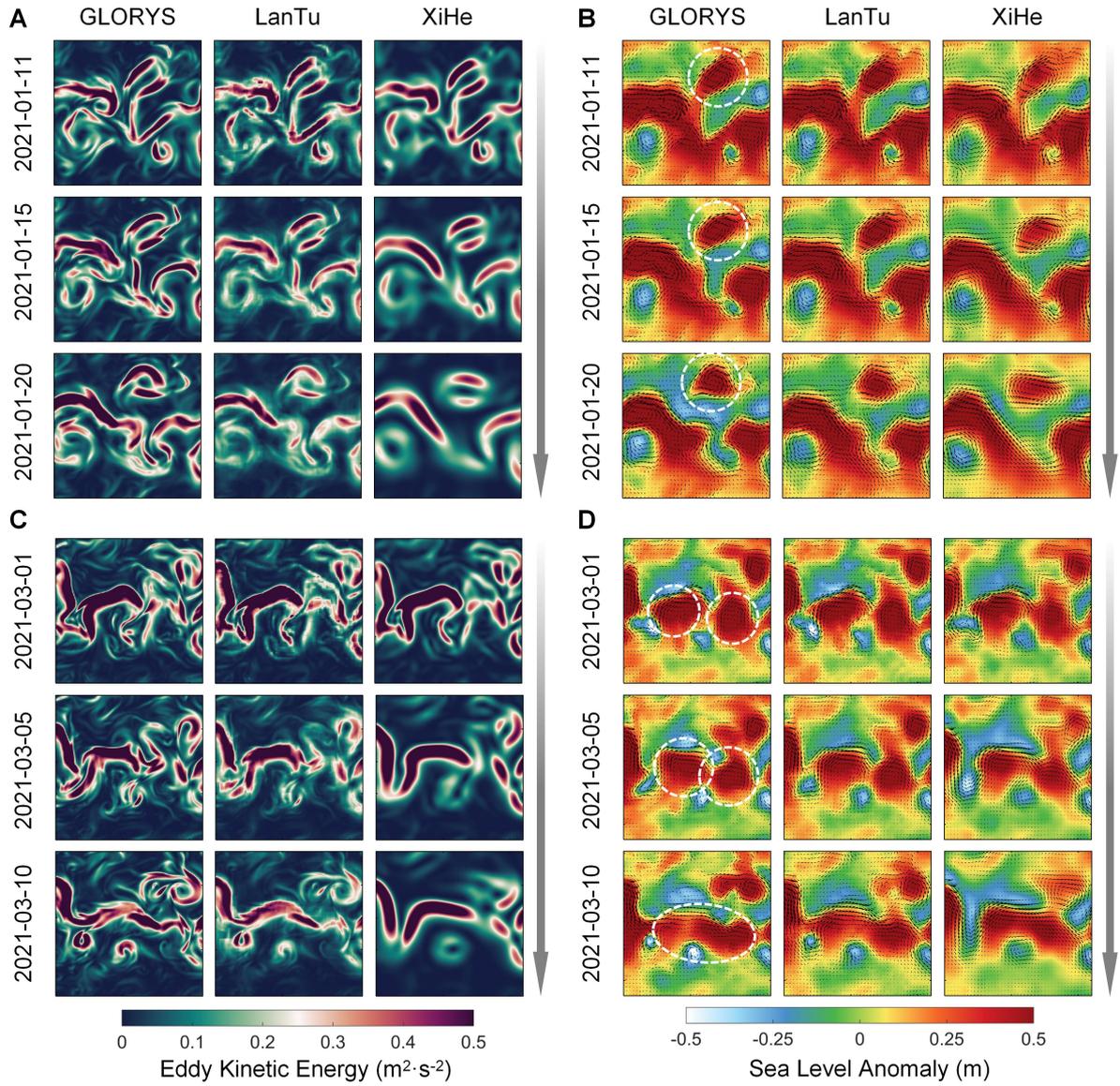

**Fig. 4.** Comparison of the differences between LanTu and XiHe in mesoscale eddy dynamics (splitting and merging) forecasting. (**A** and **B**) EKE (left) and SLA (right) of eddy splitting (initial time: 10 January 2021). (**C** and **D**) EKE (left) and SLA (right) of eddy merging (initial time: 28 February 2021).

In fact, the smoothing (or blurring) effect is commonly observed in AI forecasting models. On one hand, to meet the generalization of global ocean forecasting, AI-OFS tend to produce more robust smoothed solutions, thereby overlooking fine local features. On the other hand, AI models optimized with norm-based loss functions (such as mean squared error) tend to produce statistical averages of the labels (*31*). In developing the LanTu model, we design a dynamics-enhanced multiscale constraint to capture the evolution of mesoscale eddies (see Materials and Methods for details). In short, we incorporate the spatial correlation of forecast increments into the optimization of LanTu, rather than just relying on a norm constraint.

Learning mesoscale eddy dynamics is crucial for AI-OFS to achieve eddy-resolving ocean forecasting. If the fundamental evolution of mesoscale eddies can be captured by LanTu (Fig. 3), then forecasting the splitting and merging of eddies should be a reasonable expectation. In Fig. 4, we show two samples of LanTu forecasts near the Kuroshio in Japan, which represent eddy splitting (Fig. 4B) and merging (Fig. 4D). From the SLA, it can be seen that the spatial patterns of mesoscale eddies forecasted by LanTu and XiHe are similar. In localised details, XiHe is still smoother than LanTu. The comparison shows that there is a significant difference in EKE between LanTu and XiHe (Fig. 4A and C). The EKE of LanTu is closer to the GLORYS reanalysis, with only a slight suppression in intensity. The spatial pattern of the EKE predicted by the XiHe model undergoes significant smoothing and deformation. The above results show that the smoothing effect cannot be completely eliminated, but can be mitigated by improving the learning strategy of AI-OFS. The target region of LanTu is the Northwestern Pacific, where the evolution of mesoscale vortices is significant (*32*). Therefore, the dynamics-enhanced multiscale constraint can help LanTu learn the mesoscale processes in this region more efficiently.

## 2.3. Capture the Fine 3D Eddy Structure

The vertical structure of mesoscale eddies is key to eddy-induced heat/matter transfer, which can further affect climate and ecosystems (*12*). In the previous section, LanTu has shown excellent ability in forecasting mesoscale eddy dynamics. To further evaluate the eddy forecasting performance of LanTu, we compare the 3D eddy structures predicted by LanTu and XiHe at different lead times (Fig. 5).

We can identify a distinct cyclonic eddy from the satellite-observed SLA. Both LanTu and XiHe forecast results have reproduced the 3D eddy structure of the upper ocean. The cyclonic eddy induces a vertical motion of seawater from the bottom to surface, where cold water from the lower layers is brought to the upper layers, which leads to cooler temperatures inside the eddy (*33*). The thermodynamic and dynamic relationships predicted by LanTu are highly consistent, and it outperforms XiHe in both the spatial distribution and strength of the eddy. Although the filamentous structure of the eddy may be smoothed, LanTu still captures the key features and structure, with salinity patterns being similar as well (Fig. S5). In contrast, the XiHe shows a significant blurring of eddy details in forecasts with a lead time of 5 days.

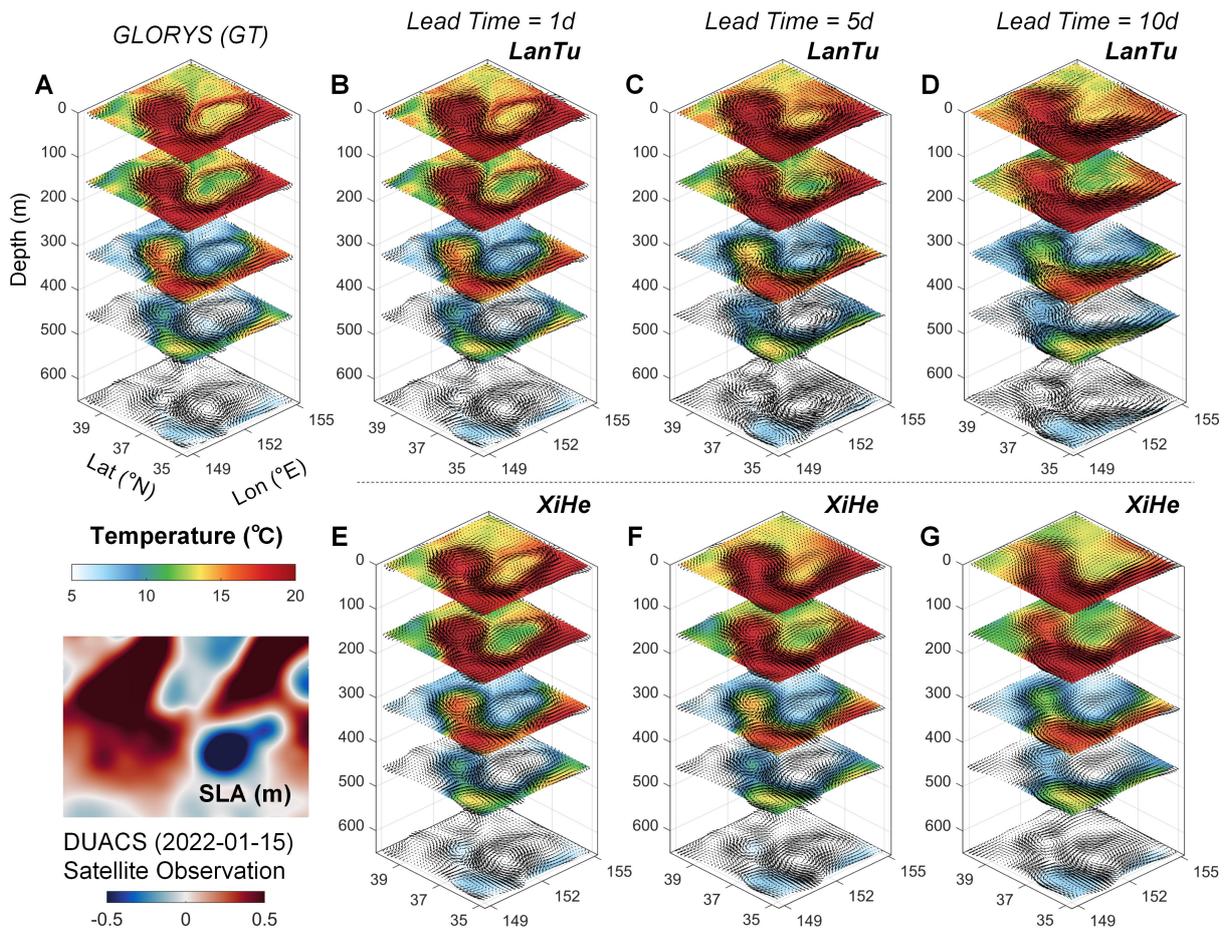

**Fig.5.** The vertical structure of the mesoscale eddy with the corresponding 3D temperature distribution. The lower left panel shows satellite-observed SLA. (**A**) Ground truth on 15 January 2022 (GLORYS), (**B** to **D**) LanTu and (**E** to **G**) XiHe. The maximum depth is 643m.

The results once again demonstrate that the LanTu model has significant advantages and potential in handling complex eddy evolution and fine structure prediction. Therefore, LanTu can effectively supplement or serve as a lightweight alternative to the regional eddy-resolving NOFS, which holds reference value for improving NOFS under physical frameworks.

## 3. Discussion

In this study, we present LanTu, an AI-based regional eddy-resolving ocean forecasting system (AI-OFS). Inspired by the physical knowledge of ocean-atmosphere interactions and multiscale dynamics, an innovative dynamic-enhanced constraint is designed for targeted optimization of LanTu, aiming to enhance its focus on mesoscale processes. The results show that the LanTu obtains significant improvement in forecasting the dynamics and 3D structure of mesoscale eddies. Compared with the existing AI-OFS, LanTu effectively mitigates issues such as the blurring or loss of eddy features in the forecasts. Compared to the existing short-term numerical forecasting benchmark (IV-TT), LanTu performs well in forecasting all ocean variables (1-day to 10-day lead times). It is worth mentioning that LanTu consistently outperforms traditional baselines (Fig. S6 to S8) in forecasting beyond 10 days (2~4 weeks).

Although the performance of LanTu is satisfactory, it still has limitations. In addition to optimising model architecture, integrating richer physics knowledge or constraints into AI models is promising for future research. Whether the importance of boundary information for AI-OFS in regional ocean modelling is consistent with our experience (mainly from NOFS) remains open to discussion. We must also emphasize that, at least for now, AI-OFS should not be viewed as a replacement for ocean numerical forecasting. Reanalysis products remain the main source for training AI-OFS, even though the quality of ocean reanalysis may not be as high as atmospheric reanalysis (*34*). Therefore, AI-OFS may be an extended application built upon the foundation of numerical models and data assimilation techniques.

LanTu has unleashed the potential of AI-OFS in the application of eddy-resolving ocean forecasting. In underdeveloped countries with limited computing or development resources, lightweight AI-OFS is more hardware-friendly for operational forecasting. Without relying on complex physical frameworks, the forecasting skill of the LanTu can be similar to NOFS, providing a reference for further research on mesoscale eddy dynamics. In addition, LanTu can be nested or coupled with other global AI-OFS or AI-based weather forecasting models, which can provide more intelligent forecasting services for humans around the world. With the rapid development of AI, we are optimistic about the future integration of data-driven and physics-driven approaches to achieve better performance.

## 4. Materials and Methods

### 4.1. GLORYS and ERA5 Reanalysis

To fully train the LanTu model, we have employed the Copernicus Global 1/12° Oceanic and Sea Ice GLORYS12 reanalysis (referred to as GLORYS). The GLORYS reanalysis system (*35*), designed and implemented in the framework of the Copernicus Marine Environment Monitoring Service (CMEMS), integrates ocean and sea ice components based on the advanced Nucleus for European Modelling of the Ocean (NEMO) framework. The GLORYS system has a horizontal resolution of 1/12°, with 50 vertical layers (denser in the upper layers and sparser in the lower layers). The reduced-order Kalman filter is employed to assimilate observations into the model, including in-situ temperature and salinity profiles, as well as satellite-based SLA, sea surface temperature (SST) and sea ice concentration (SIC). The ERA-Interim reanalysis provided by the European Centre for Medium-Range Weather Forecasts (ECMWF) serves as atmospheric forcing for NEMO. The daily GLORYS data from January 1, 1993, to December 31, 2020, is employed for training the LanTu and the data from 2021 to 2023 is used for evaluation.

ERA5 is the fifth generation of the ECMWF reanalysis dataset (*36*). ERA5 provides hourly surface and upper-air atmospheric data with a spatial resolution of 0.25° from 1940 to the present. The accuracy of ERA5 has been widely verified. We employ mean sea level pressure (SLP), 2m air temperature (T2M), 10m wind u component (U10) and 10m wind v component (V10) as atmospheric drivers for LanTu. To unify the spatial and temporal resolution, we regrid the atmospheric variables from ERA5 onto the GLORYS grid and convert the hourly data into daily averages.

### 4.2. LanTu Ocean Model

#### 4.2.1 Overall Architecture

The LanTu model focuses on forecasting the daily averaged state variables within the upper 643m of the ocean. In addition to SLA, LanTu operates in 23 vertical depth levels (Table S3), with each level containing four forecast variables: temperature, salinity, zonal current and meridional current. Different ocean variables are stacked according to their depth levels, with variables at the same depth level being prioritized for merging. The four surface atmospheric variables are placed above all the ocean variables. The LanTu input data combines surface atmospheric conditions and ocean state variables to generate a data cube with dimensions of 97×780×1080. Here, 97 represents the total number of input variables, 780 corresponds to the grid points for latitude (H) and 1080 corresponds to the grid points for longitude (W). All the data need to be standardized before training.

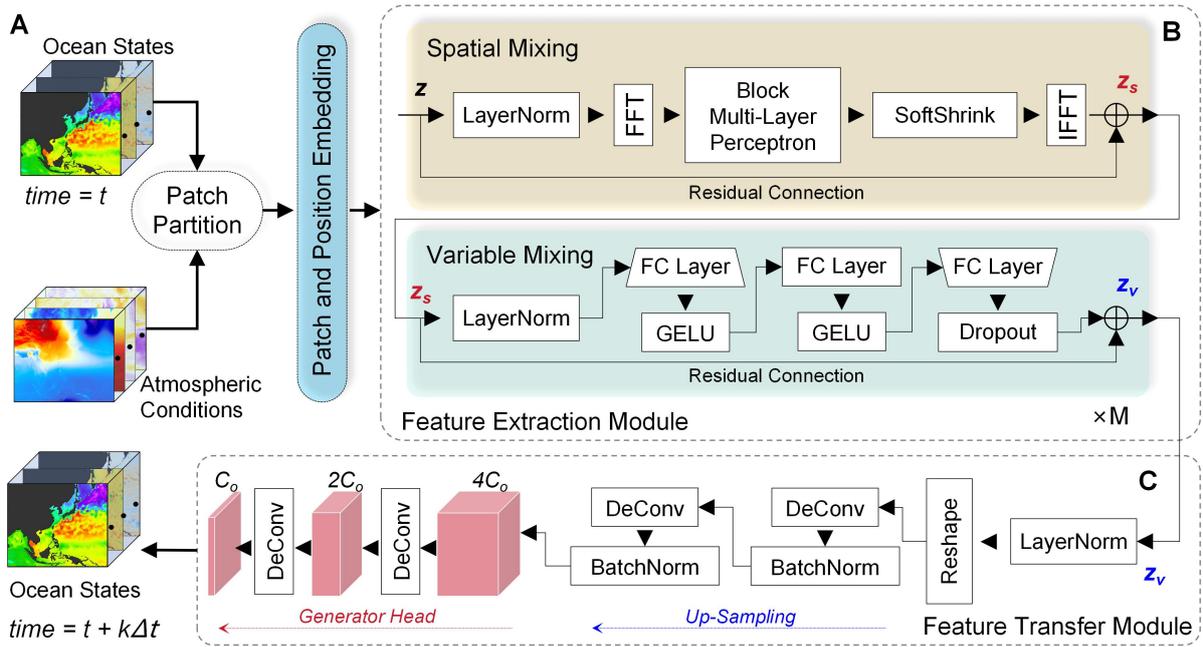

**Fig.6.** Overall structure and design details of LanTu. (**A**) Feature embedding, which maps the initial fields to a latent space. (**B**) Feature extraction module, which is used to extract and fuse multiscale features. (**C**) Feature transfer module employs cascaded transposed convolution to generate eddy-resolving ocean forecasts. LanTu adopts a non-autoregressive strategy, training a separate model for each forecast lead time.

The architecture of LanTu consists of three main components, as shown in Fig. 6: feature embedding, feature extraction and feature transfer. Considering the modeling and forecasting requirements for eddy-resolving, we refer to the architecture of the Adaptive Fourier Neural Operator (AFNO), which is specifically designed for high-resolution inputs (*37*). LanTu adopts a modeling approach that combines AFNO with Vision Transformer (VIT).

First, the high-dimensional input data is divided into evenly spaced feature grids, with each grid representing a feature patch cube. All feature patches and their positional encodings are embedded into a high-dimensional feature space with a large number of latent channels.

The embedded features represent the dedicated "semantic" encoding of ocean-atmosphere variables and geographical coordinates (Fig. 6A).

Next, the embedded features are processed by a Feature Extraction Module (FEM). The internally cascaded FEM performs continuous spatial mixing and variable mixing to extract the multiscale nonlinear features (Fig. 6B).

Finally, the transposed convolution decoder in the Feature Transfer Module (FTM) decodes high-dimensional latent features into future ocean states (Fig. 6C).

### 4.2.2 Details of the Model Components

In this study, the feature patch cube has a size of 97×4×4, and the high-dimensional input data is projected onto a 195×270 feature grid. The feature embedding is a convolutional layer (with an embedding dimension of 768). The kernel size and stride are chosen as 4 and the output is the embedded latent features, with a dimension of 52650×768.

The Feature Extraction Module (FEM) consists of two parts: the spatial mixing block and the variable mixing block, with M representing the depth of FEM, which is set to 12. The spatial mixing block employs Fourier operators to extract spatial multiscale information from the embedded features. The variable mixing block consists of a linear layer and a Gaussian Error Linear Unit (GELU), which focuses on the nonlinear dependencies between different variables. FEM employs residual connections to enhance the representation of the neural network. The Feature Transfer Module (FTM) consists of an upsampling block and a generator head with cascaded transposed convolution layers. During the upsampling, the transposed convolution layers (with a kernel size and stride of 2) reshape the output dimension of the FEM from 768×195×270 to 372×780×1080. Finally, while maintaining the spatial dimensions of the features, the generator head further compresses the output dimension to 93×780×1080 to get the final output. To ensure the spatial integrity of the variables and the focus on ocean forecasting, LanTu employs a land mask to replace the land points in ocean variables with 0.

### 4.2.3 Training Strategy

The LanTu model is developed using the PyTorch framework and is trained on a cluster consisting of 4 Nvidia A100 GPUs. To avoid cumulative errors, the LanTu model generates forecasts in a non-autoregressive strategy. We train a separate model for each forecast step:

$$\hat{\mathbf{O}}^{t+k\Delta t} = \text{LanTu}(\mathbf{O}^t, \mathbf{A}^t, \theta), k = 1, 2, \cdots, K \tag{1}$$

where $\mathbf{O}^t$ and $\mathbf{A}^t$ represent the initial fields of the ocean and atmosphere, respectively. $\theta$ represents the model parameters. The forecast step $\Delta t$ is 1 day and $K$ is the lead time. Each model is trained for 100 epochs. We employ the Adam optimizer with a learning rate set to $5 \times 10^{-5}$.

### 4.2.4. Design of the Dynamics-Enhanced Multiscale Constraint

When training the LanTu model, we design a dynamics-enhanced multiscale constraint. The classic norm constraint can capture the evolution of large-scale features, but it tends to overlook mesoscale signals, which leads to smooth results during model inference. To address this limitation, the loss function of LanTu consists of two parts: static constraint ($Loss_S$) and dynamic constraint ($Loss_D$). The static constraint uses mean squared error (MSE) to quantify the forecast error at the target moment. The dynamic constraint is the forecast increment correlation constraint based on Pearson correlation coefficient to better capture the mesoscale dynamic evolution. The goal of this design is to minimize the MSE while maximizing the correlation between the forecast increments and ground truth increments, which is defined as follows:

$$Loss = \lambda_S \times Loss_S + \lambda_D \times Loss_D \tag{2}$$

$$Loss_S = \frac{1}{H \times W \times C_{out}} \sum_{i=1}^{H} \sum_{j=1}^{W} \sum_{c=1}^{C_{out}} \left( \hat{\mathbf{O}}_{i,j,c}^{t+k\Delta t} - \mathbf{O}_{i,j,c}^{t+k\Delta t} \right)^2 \tag{3}$$

$$Loss_D = 1 - Corr\left( \Delta \mathbf{O}^{t+k\Delta t}, \Delta \hat{\mathbf{O}}^{t+k\Delta t} \right) \tag{4}$$

where $\lambda_S$ and $\lambda_D$ represent the weights of the two types of constraints, both set to 0.5. $\Delta \hat{\mathbf{O}}^{t+k\Delta t}$ and $\Delta \mathbf{O}^{t+k\Delta t}$ represent the forecast increments and ground truth increments relative to the initial field for LanTu and GLORYS, respectively. In the calculation of dynamic constraint, we have removed all land points.

### 4.3. Baseline Methods and Evaluation Datasets

We employ the IV-TT Class 4 framework (*28*) to evaluate the forecasting performance of the LanTu model, which is widely used for the quantification and comparison of operational ocean forecasting skills. IV-TT regularly publishes short-term forecasts of the mainstream global NOFS, including FOAM, PSY4, GIOPS, BLUElink OceanMAPS (BLK) and others. We employ the daily IV-TT data from 2021 to 2023 to assess the forecasting skill of the LanTu. The detailed IV-TT configuration is provided in Table S1.

In addition to numerical forecasting methods, we have adopted several statistical and AI baselines for comparison experiments. These primarily include persistence, the XiHe model and the autoregressive version of LanTu model (LTAR). Persistence is the simplest and most widely used baseline method, which assumes that the forecasts remains the same as the initial value. LTAR employs the same model structure as LanTu, but outputs both ocean and atmosphere states. LTAR uses its own output as input, allowing it to generate forecasts for different lead times. To ensure fairness in evaluation, neither LTAR nor LanTu underwent any fine-tuning. XiHe is a data-driven global AI-OFS and its details have been described earlier.

We also employ the Ssalto/Duacs altimeter products (*38*) generated and distributed by the CMEMS. The product integrates measurements from multiple satellite altimeters through optimal interpolation. We employ the daily SLA and ADT (Absolute Dynamic Topography) from 2021 to 2023 in this product.

## 4.4. Evaluation Metrics

We use the anomaly correlation coefficient (ACC), root mean square error (RMSE) and persistence skill score (PSS) to evaluate the forecasting performance. The ACC is used to measure the correlation between the forecast and ground truth, and it is commonly used to quantify the effective forecast lead time (ACC>0.6). In the calculation, we subtract the daily climatology. It is defined as:

$$ACC = \sum_{i=1}^{N} (\mathbf{O}_i - \mathbf{C}_i)(\hat{\mathbf{O}}_i - \mathbf{C}_i) \bigg/ \left[ \sqrt{\sum_{i=1}^{N} (\mathbf{O}_i - \mathbf{C}_i)^2} \sqrt{\sum_{i=1}^{N} (\hat{\mathbf{O}}_i - \mathbf{C}_i)^2} \right]$$ (5)

where $\mathbf{O}$ and $\hat{\mathbf{O}}$ represent the ground truth and forecasts, respectively, $\mathbf{C}$ represents the corresponding daily climatology. $N$ represents the total number of water grid points in the ocean state. RMSE is an error metric used to measure the distance between the forecast and the ground truth, effectively reflecting the accuracy of forecasts. The smaller the RMSE, the better the forecasting performance of the model. It is defined as:

$$RMSE = \sqrt{\frac{1}{N}\sum_{i=1}^{N}\left(\mathbf{O}_i - \hat{\mathbf{O}}_i\right)^2} \tag{6}$$

PSS is used to evaluate the improvement of a model relative to persistence forecast. It is defined as:

$$PSS = 1 - \frac{RMSE_L}{RMSE_P} \tag{7}$$

where $RMSE_L$ and $RMSE_P$ represent the RMSE of the LanTu model and the persistence, respectively.

## Data availability

The GLORYS reanalysis datasets are obtained from https://data.marine.copernicus.eu/product/GLOBAL_MULTIYEAR_PHY_001_030/description. The ERA5 reanalysis datasets are obtained from https://cds.climate.copernicus.eu/datasets/reanalysis-era5-single-levels?tab=overview. The SLA observations are obtained from https://data.marine.copernicus.eu/product/SEALEVEL_GLO_PHY_L4_MY_008_047/description. The IV-TT Class 4 framework: https://thredds.nci.org.au/thredds/catalog/rr6/intercomparison_files/catalog.html. The inference codes and model weights of XiHe can be found at https://github.com/Ocean-Intelligent-Forecasting/XiHe-GlobalOceanForecasting.

## Acknowledgements

This work was supported in part by the National Natural Science Foundation under Grants 42376190, 41876014 and 42406191, in part by the National Key Research and Development Program under Grants 2023YFC3107800, 2022YFC3104800 and 2021YFC3101500.

Large-Scale Climate: A Review. *Journal of Climate* **36**, 1981–2013 (2023).

Supplementary Materials for

# LanTu: Dynamics-Enhanced Deep Learning for Eddy-Resolving Ocean Forecasting

Qingyu Zheng[a], Qi Shao[b], Guijun Han[a], Wei Li[a,*], Hong Li[a,*], Xuan Wang[a]

[a] *Tianjin Key Laboratory for Marine Environmental Research and Service, School of Marine Science and Technology, Tianjin University, Tianjin 300072, China*

[b] *Fujian Key Laboratory on Conservation and Sustainable Utilization of Marine Biodiversity, Minjiang University, Fuzhou 350108, China*

\* Corresponding author. *E-mail:* liwei1978@tju.edu.cn (W. Li), hongli@tju.edu.cn (H. Li).

**This PDF file includes:**

    Supplementary Text

    Figs. S1 to S8

    Tables S1 to S3

**Table S1.** Details of the ocean forecasting systems used for comparison in this study

| System | Forecast range | Number of vertical levels | Ocean Model | Horizontal resolution | Spatial range |
|--------|---------------|--------------------------|-------------|----------------------|---------------|
| BLK | 7 days | 51 | MOM 5 | 1/10° | Global |
| FOAM12 | 6 days | 75 | NEMO 3.2 | 1/12° | Global |
| FOAM25 | 6 days | 75 | NEMO 3.2 | 1/4° | Global |
| GLOPS | 10 days | 50 | NEMO 3.1 | 1/4° | Global |
| PSY4 | 10 days | 50 | NEMO 3.1 | 1/12° | Global |
| XiHe | 10 days | 23 | Deep learning | 1/12° | Global |
| LanTu | 30 days | 23 | Deep learning | 1/12° | Regional |

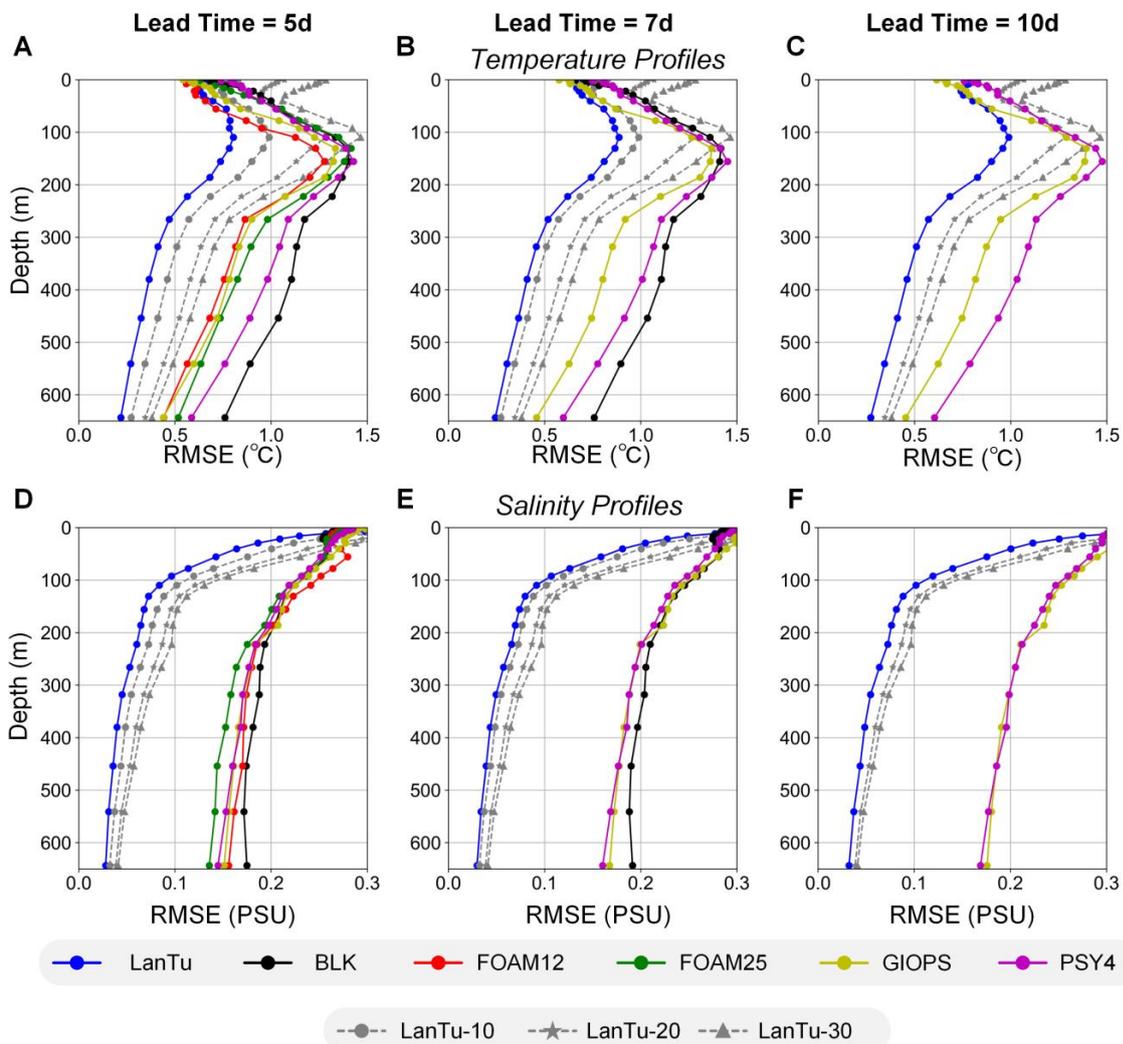

**Fig. S1.** Comparison of root mean square error for NOFS temperature (**A** to **C**) and salinity (**D** to **F**) profile forecasts for LanTu and IV-TT. Forecast lead times of 5, 7 and 10 days. The grey dashed lines represent forecasts with 10-day (LanTu-10, circle), 20-day (LanTu-20, pentagram) and 30-day (LanTu-30, triangle) lead times for LanTu.

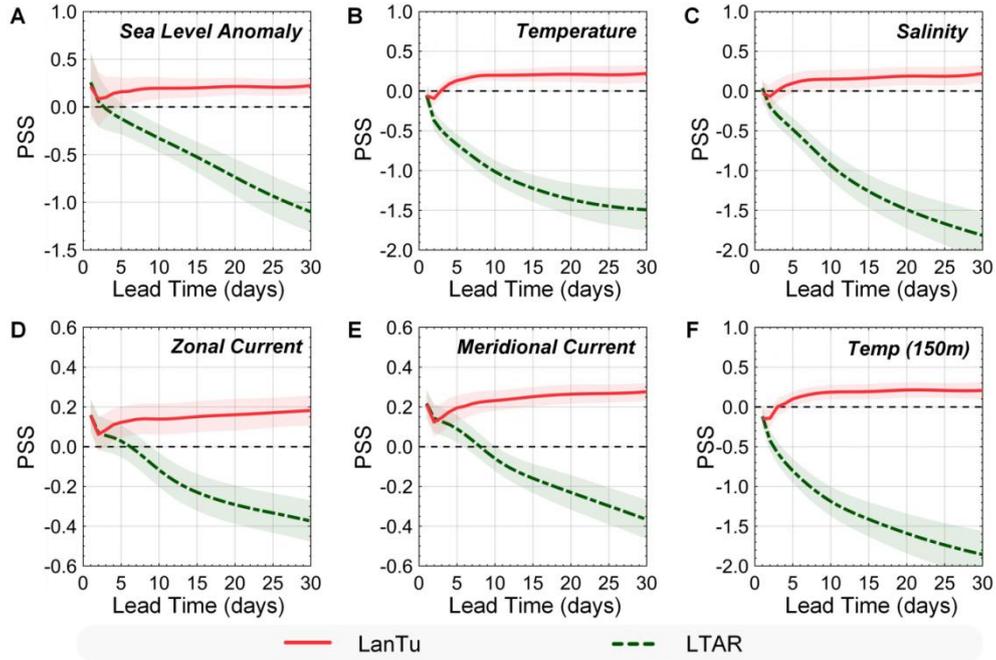

**Fig. S2.** Comparison of Persistence Skill Score (PSS) between LanTu and LTAR. PSS (higher is better) for forecast (**A**) SLA, (**B**) temperature, (**C**) salinity, (**D**) zonal current, (**E**) meridional current and (**F**) 150-m temperature versus forecast lead time. The zero-lead time represents the initial conditions. Red and green lines correspond to LanTu and LTAR, respectively.

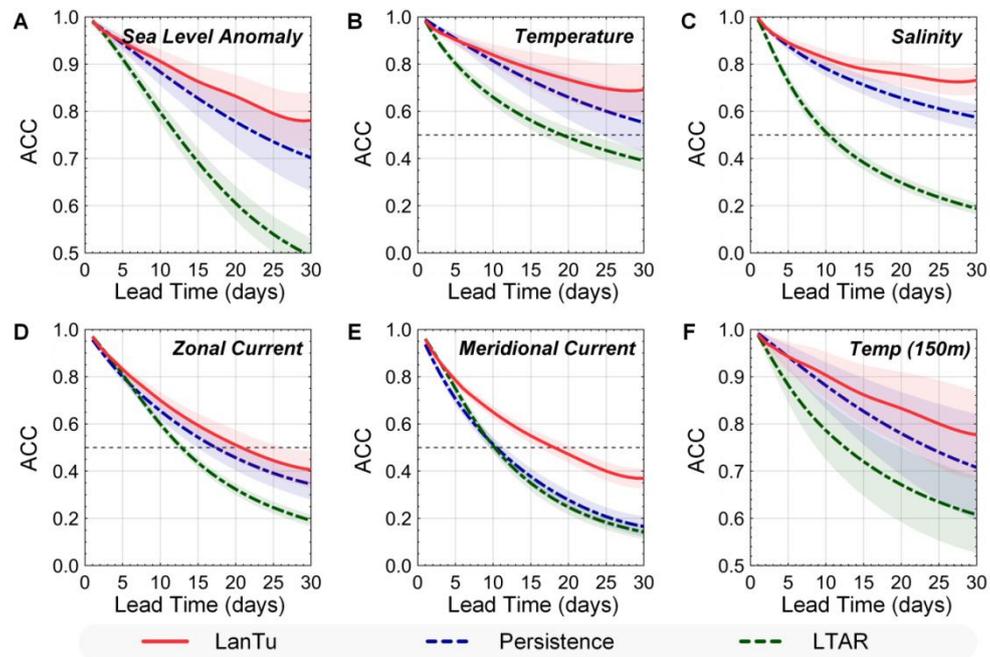

**Fig. S3.** Comparison of anomaly correlation coefficients (ACC) between LanTu, LTAR and persistence. ACC (higher is better) for forecast (**A**) SLA, (**B**) temperature, (**C**) salinity, (**D**) zonal current, (**E**) meridional current and (**F**) 150-m temperature versus forecast lead time. The zero-lead time represents the initial conditions. Red and green lines correspond to LanTu and LTAR, respectively.

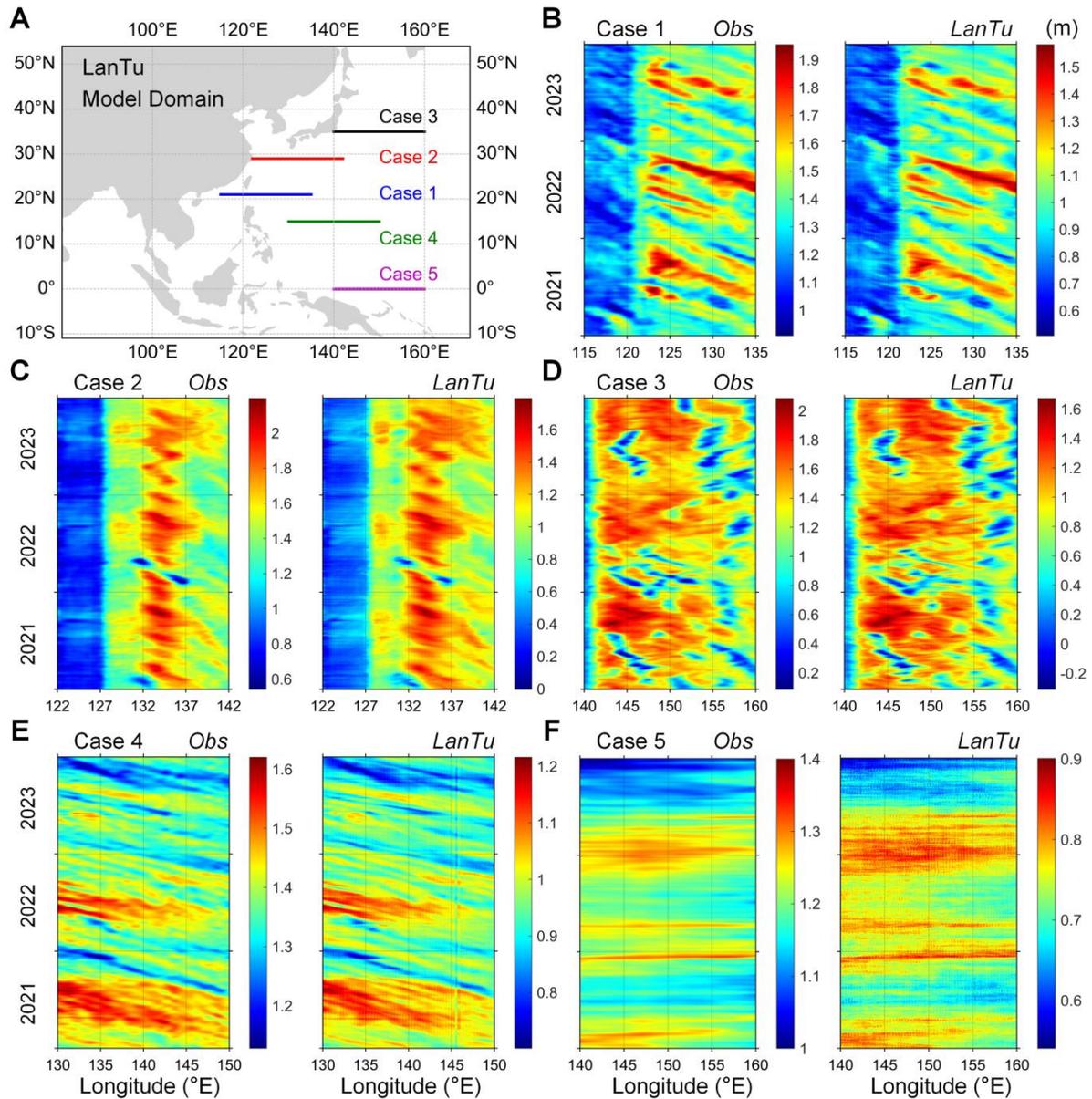

**Fig. S4.** Compare the Hovmöller diagrams of the Sea Surface Height (SSH) between LanTu and satellite observations. (**A**) The model domain of LanTu and the positions of the five zonal sampling bands. (**B-F**) Case1 to Case5 for the Hofmöller diagram from 1 January 2021 to 31 December 2023. Satellite observations in the left panel and LanTu in the right panel.

**Table S2.** Latitude and longitude information for Case 1 to Case 5 used for plotting the Hofmöller diagram.

| Coordinates | Case 1 | Case 2 | Case 3 | Case 4 | Case 5 |
|---|---|---|---|---|---|
| Latitude | 21°N | 29°N | 35°N | 15°N | 0 |
| Longitude | 115~135°E | 122~142°E | 140~160°E | 130~150°E | 140~160°E |

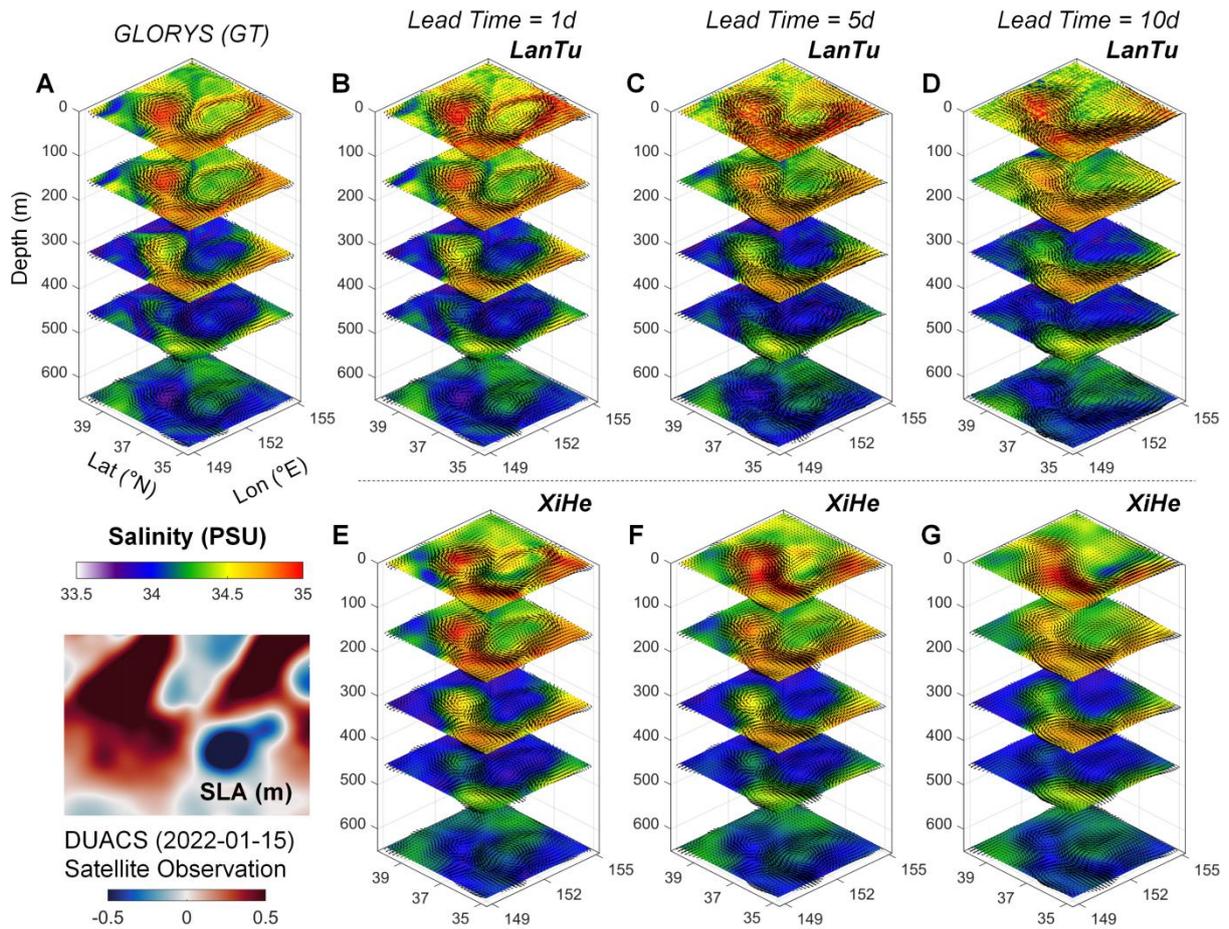

**Fig. S5.** The vertical structure of the mesoscale eddy with the corresponding 3D salinity distribution. The lower left panel shows satellite-observed SLA. (**A**) Ground truth on 15 January 2022 (GLORYS), (**B** to **D**) LanTu and (**E** to **G**) XiHe. The maximum depth is 643m.

**Table S3.** Vertical 23 layers of the LanTu model with corresponding depths.

| Level | 1 | 2 | 3 | 4 | 5 | 6 | 7 |
|-------|------|------|------|------|------|------|------|
| Depth | 0.49m | 2.65m | 5.08m | 7.93m | 11.41m | 15.81m | 21.60m |
| Level | 8 | 9 | 10 | 11 | 12 | 13 | 14 |
| Depth | 29.44m | 40.34m | 55.76m | 77.85m | 92.32m | 109.73m | 130.67m |
| Level | 15 | 16 | 17 | 18 | 19 | 20 | 21 |
| Depth | 155.85m | 186.13m | 222.48m | 266.04m | 318.13m | 380.21m | 453.94m |
| Level | 22 | 23 | | | | | |
| Depth | 541.09m | 643.57m | | | | | |

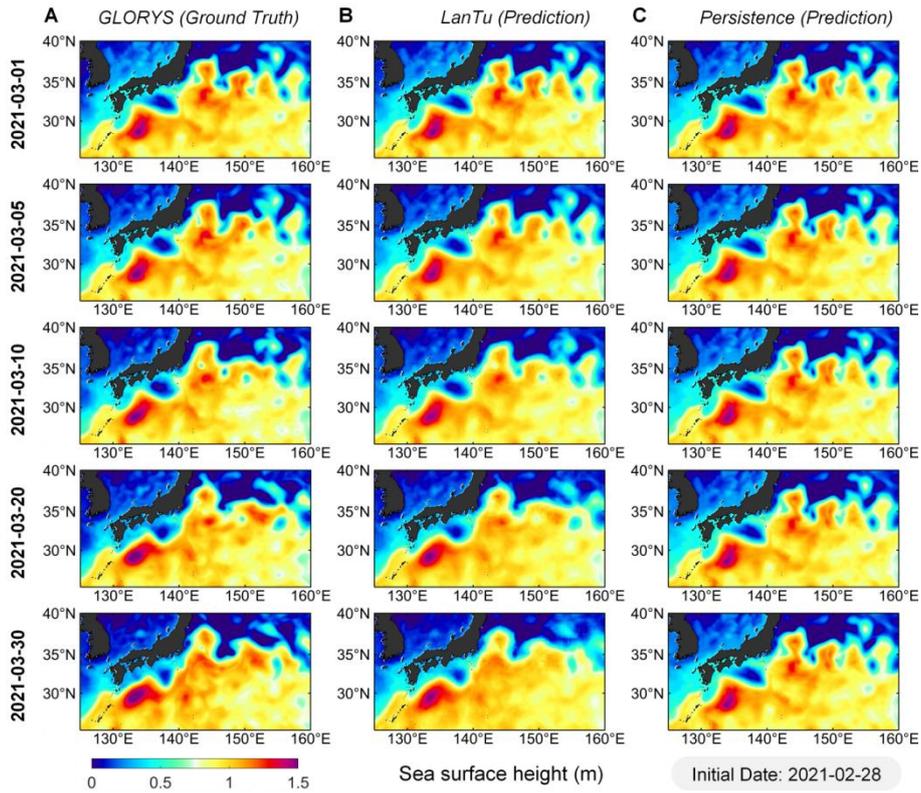

**Fig. S6.** Forecast visualisation of sea surface height (SSH) from 1 March to 30 March 2021 (**A**) GLORYS, (**B**) LanTu and (**C**) Persistence.

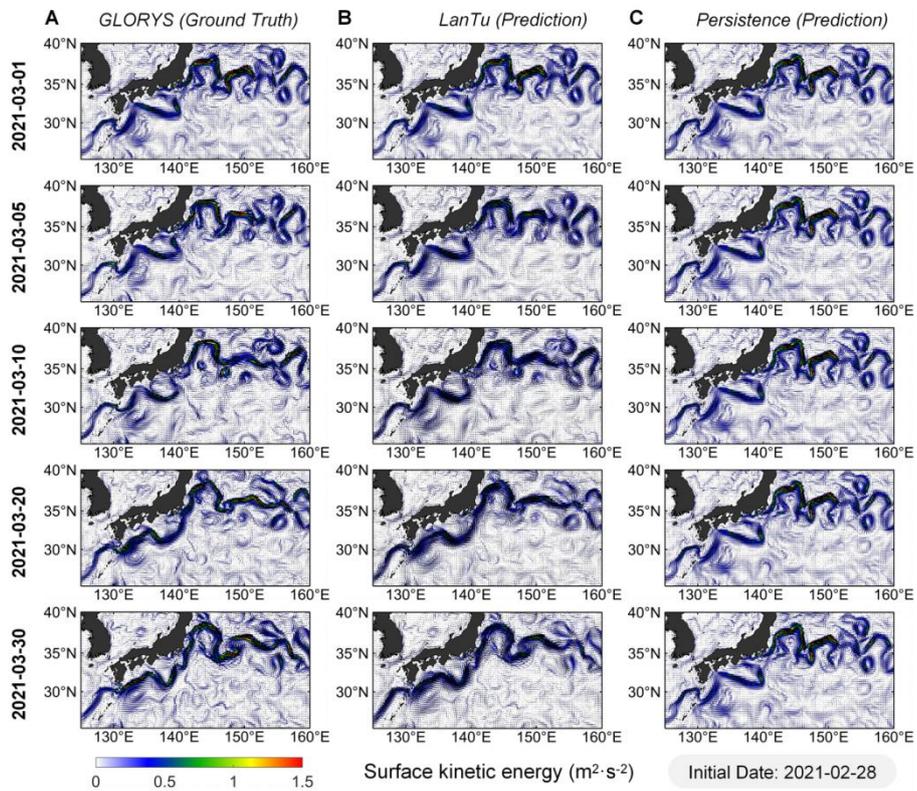

**Fig. S7.** Same as Fig. S6, but for eddy kinetic energy (EKE).

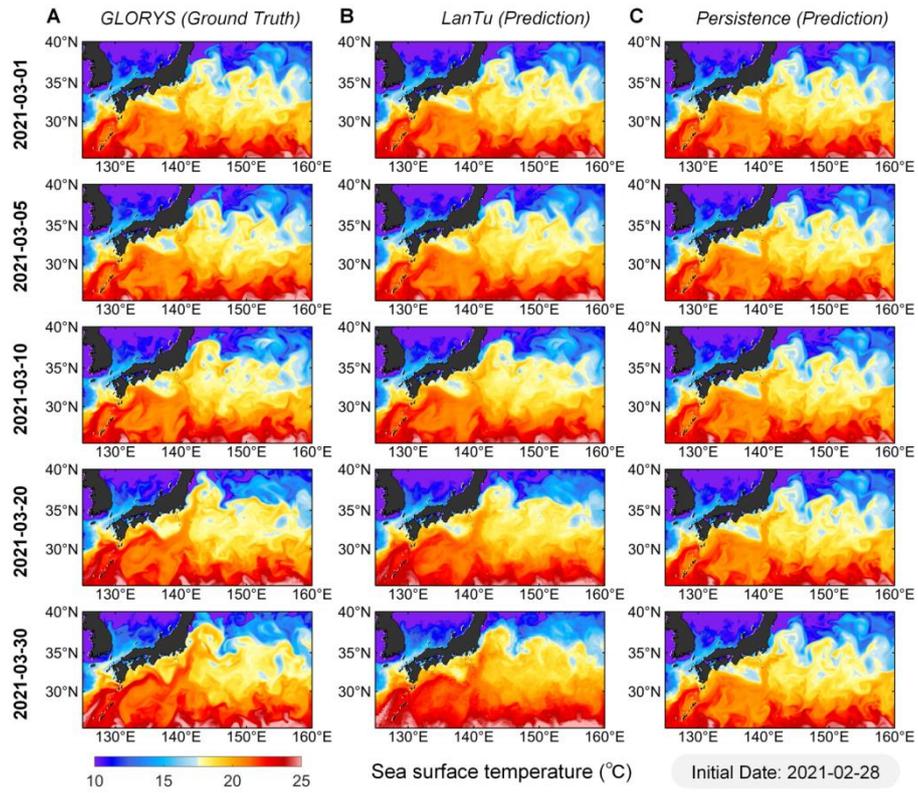

**Fig. S8.** Same as Fig. S6, but for sea surface temperature (SST).